# A novel direct structured-light inspection technique for contaminant and defect detection

## Yiyang Huang[*]

*Corresponding author: yiyanghuang_acad@hotmail.com

The Direct Structured-Light Inspection Technique (DSIT) proposed in this paper is a novel method that can be implemented under two types of binary structured light illumination to detect contaminants and defects on specular surfaces and transparent objects, in which light reflection system is used to detect specular surfaces, while light transmission system is applied for transparent object inspection. Based on this technique, contaminant and defect distribution can be directly obtained without any calculation process. Relevant simulations and experiments are performed to prove the effectiveness of DSIT.

## 1. INTRODUCTION

Quality control of optical components has become one of the most important subjects in manufacturing industries. Contaminants and defects on optical components can greatly affect the performance of the device equipped with these components, sometimes can even bring about disasterous subsequences [1]. Limited by the detection accuracy and speed, optical component inspection in industrial environments is currently dominated by human visual approach. To improve this situation, a variety of detection methods based on machine vision have been proposed for automatically detecting contaminants and defects of optical components. According to the difference in system setup, these inspection methods can be mainly classfied into several types: dark-field imaging method, diffuse backlight method, multi-angle lighting method, and so on [2].

In recent years, detection methods based on structured light have developed rapidly [3-6], in which one of the new methods proposed for contaminant and defect detection is called Structured-Light Modulation Analysis Technique (SMAT) [7, 8]. SMAT is a new approach that employs the modulation image as its detection basis, which is suitable for quality inpection of specular surfaces and transparent objects in industrial environments.

Inspired by the optical mechanism of contaminants and defects in Refs. [7, 8], in this paper, a novel Direct Structured-Light Inspection Technique (DSIT) based on binary structured light illumination is proposed for contaminant and defect detection, and this new technique can be implemented by means of a light reflection system or a transmission system. DSIT is applicable for the detection of many different kinds of contaminants and defects on specular surfaces or transparent objects, such as dust, scratches, cracks, and breakages, etc. In the following, Section 2 contains the detailed mechanism analyse of DSIT. The simulation and experimental results are placed in Section 3 and Section 4, respectively. Section 5 is the summary of this work.

It should be noted that at first glance, the mechanism of DSIT may be mistaken for the same as dark-field method or diffuse backlight method, but in fact their detection mechanisms are completely different. First, the dark-field method adopts a lateral light source to supply illumination for the detection system. It does not provide incident light to most areas of the detected object, but only illuminate contaminants and defects. DSIT aims to provide whole-field illumination for all parts of the detected object. Second, the light source used in the dark-field imaging system is usually ordinary light sources, while DSIT adopts planar structured light for its illumination. Third, the detection light used in the dark-field imaging method is essentially reflected light, while DSIT can use not only reflected light, but also transmitted light. Compared with the diffuse backlight method, the DSIT based on transmission system uses the binary structured light specially designed in this work, instead of the uniform illumination adopted by diffuse backlight method.

## 2. PRINCIPLE

*2.1. The relations of light rays*

The detection methods based on structured light illumination can be applied by means of a reflection system or a transmission system, and the system structures are shown in Fig. 1. The display screen is used for projection of structured light, and the light patterns reflected or transmitted by the detected object will be collected by the camera and then processed by the computer.

The core mechanism for the detection under structured light is the difference of ray relations between the places where contaminants or defects locate and the places where are clean and intact. When there is a contaminant or defect on the detected object, the topography of the detected surface will be skewed [9], which will lead to a change in light relation on it. Moreover, contaminants can sometimes lead to changes in the reflectivity of their locations.

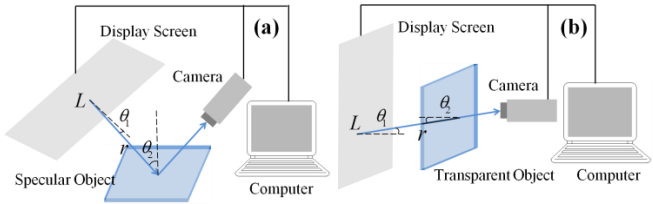

Fig. 1. The structure of (a) the reflection system, (b) the transmission system.

Based on the optical path reversible theorem, when the reflected light path and transmitted light path are determined, the change in the surface topography of objects will bring about a transformation in incident light rays, which can be observed in Fig. 2. The incident light rays in Figs. 2(a)-2(b) are derived based on the determined reflected light and transmitted light collected by one pixel of camera imaging target. From Fig. 2, it can be clearly seen that the original incident light source regions are deviated and diverged to larger regions.

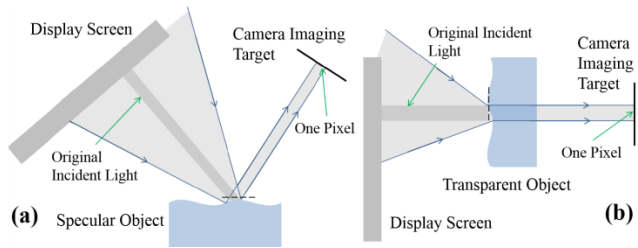

Fig. 2. The light relation in (a) the reflection system, (b) the transmission system.

### 2.2. Mathematical model for the detection based on structured light

The conversion in incident light will lead to changes in the brightness of one pixel in camera imaging target, which can be verified by captured images. The luminous flux in photometry is an appropriate parameter for quantifying the brightness of various sites of the detected object on captured images. To get it, we need to use some common photometric equations.

The display screen used for illumination is a Lambertian radiator with the characteristic of anisotropic illumination. The illuminance of the planar light source at a distance $r$ is [10]:

$$E = \frac{L dA_s \cos\theta_1 \cos\theta_2}{r^2} \quad (1)$$

$L$ represents the luminance of the planar light source. $\theta_1$ is the solid angle between the normal line of the light source and the light ray emitted by this source region. $\theta_2$ is the angle between the normal line of the illuminated plane and the light ray from the planar light source. $dA_s$ is the area of the planar light source region.

As can be seen from Fig. 2, the presence of contaminants and defects makes the composition of incident light more complicated. For better analyzing the luminous flux of contaminants and defects, the region on the detected surface captured by one pixel of camera imaging target is divided into many smaller flat regions (Elements), and it is assumed that the incident light in each Element comes from a constant light source. The geometric relation between the light source and the illuminated Element whose incident light comes from the display screen is shown in Fig. 1. Eq. (1) can be used as the theoretical basis for analyzing these Elements.

Let the illuminance from environment be $C(x, y)$. According to Eq. (1) and Fig. 1, the luminous flux in one pixel of camera imaging target can be deduced:

$$\Phi = \alpha \left( \sum_{d=d_0}^{D} s^{(d)} \frac{L^{(d)} A_s}{\left(r^{(d)}\right)^2} \cos\theta_1^{(d)} \cos\theta_2^{(d)} + \sum_{e=e_0}^{E} s^{(e)} C(x, y) \right) \quad (2)$$

$A_s$ is the area of the constant light source on the display screen. The superscript $(d)$ represents the Elements whose incident light comes from the display screen, and the superscript $(e)$ indicates the Elements whose incident light comes from environment. $s$ represents the area of Elements. $\alpha$ can indicate the surface reflectivity of specular surfaces and the transmissivity of transparent objects.

According to specific structured light and concrete system setting, the luminous flux in one pixel of camera imaging target can be obtained by using Eq. (2). More details about the calculation process can refer to Ref. [8].

### 2.3. Proposed binary structured light

In Refs. [7, 8], a modulation detection method based on fringe patterns and a direct inspection method adopting uniform illumination have been discussed. It can be concluded that if the calculation process of modulation can be avoided and filming time can be shortened [8], and if the issue of invisibility of contaminants and defects in the uniform illumination system can be solved [7], the detection approach adopting structured light illumination will become faster and more effective.

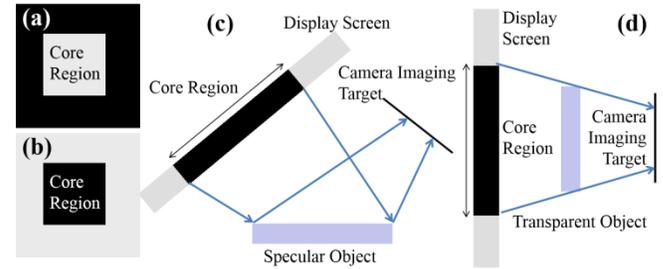

Fig. 3. The proposed (a) type 1, (b) type 2 binary structured light. The relation between the detected object and the structured light in (c) the reflection system, (d) the transmission system.

To address these needs, two types of binary structured light are specially designed in this work, and the light patterns are shown in Figs. 3(a) and 3(b). Meanwhile, the spatial relations between the detected object and structured light are shown in Figs. 3(c) and 3(d).

When the detected object is clean and intact, the light collected on the detected surface will come from the core region of the structured light. When there is a contaminant or a defect on the detected object, according to Fig. 2, it can be known that the light captured at where the contaminant or defect is situated is likely to contain the light from outside the core region, which may lead to a change in luminous flux. By virtue of the difference in luminous flux between the clean and intact places and the sites where contaminants and defects are located, the purpose of detecting contaminants and defects can be achieved under the type 1 and type 2 binary structured light.

## 3. SIMULATION

*3.1. Simulation instruction*

In the following, simulations of contaminants and defects on the detected object are performed from four items for better comparison:
(1) Modulation detection based on fringe structured light [8].
(2) Direct detection adopting uniform illumination [7].
(3) Direct detection based on type 1 binary structured light.
(4) Direct detection based on type 2 binary structured light.

It is assumed that the region on the detected surface captured by one pixel of camera imaging target is as shown in Fig. 4(a) [9]. This figure shows the topography of a punctate defect, which can be expressed as:

$$T = 100 - \frac{w}{2}\left(\sin\frac{\pi x}{100} + \sin\frac{\pi y}{100}\right) \quad (3)$$

$w$ is the parameter that controls the overall curvature of the surface topography.

The procedure of calculating simulation results based on above items (1)-(4) is as follows:

(a) Calculate the output response in one pixel as $w$ increases. The modulation response in item (1) can be calculated with the help of Ref. [8], which will not be described in detail here. The luminous flux response in items (2)-(4) can be computed according to different structured light and Eq. (2), and the detailed calculation process of Eq. (2) can also refer to the derivation process of luminous flux in Ref. [8].

(b) Calculate the output values of items (1)-(4) when the 100*100 region in Fig. 4(a) is uniform and flat, and these values can be regarded as the original outputs when the detected object is clean and intact.

(c) Divide the responses in step (a) by the values in step (b) to obtain relative results. These normalized relative results are more effective in illustrating the contrast of contaminants and defects versus ordinary areas on the obtained images.

*3.2. Results and analyses for simulation*

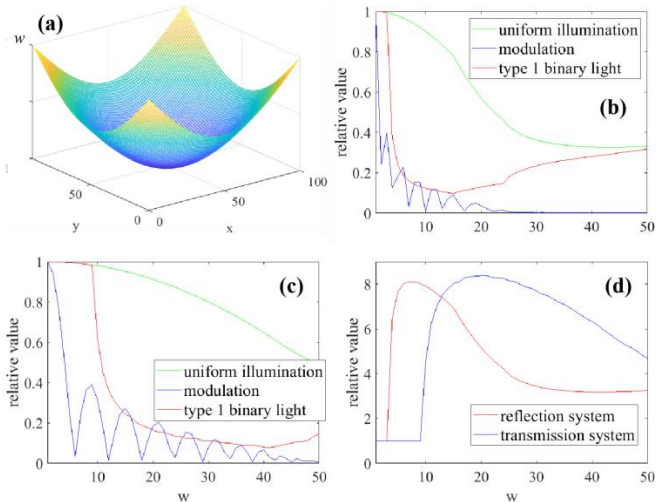

Fig. 4. (a) The topography of the detected object captured by one pixel of camera imaging target. The comparison of relative modulation, relative luminous flux based on type 1 binary structured light and that based on uniform illumination in (b) the reflection system, (c) the transmission system. (d) The relative luminous flux under type 2 binary structured light.

Based on steps (a)-(c), the relative results of items (1)-(4) can be obtained by means of a reflection system or a transmission system.

Due to the consistency of normalized range, the relative modulation, relative luminous flux under uniform illumination and that under type 1 binary structured light are settled in a single figure, and the results in the reflection system and transmission system are shown in Figs. 4(b) and 4(c), respectively. As for the luminous flux under type 2 binary structured light, the relative results obtained based on the reflection system and transmission system are plotted in Fig. 4(d).

From Figs. 4(b) and 4(c), it can be discovered that the appearance of contaminants and defects will lead to a value decreasing at where they are situated. The decline of relative luminous flux based on type 1 binary structured light is stronger than that based on uniform illumination, and weaker than the decline of modulation, which means that the result under type 1 binary structured light is better than the result employing uniform illumination, and worse than the modulation result. From Fig. 4(d), it can be found that under type 2 binary structured light, the presence of contaminants and defects will bring about an increasing in relative luminous flux values at their locations.

## 4. EXPERIMENT

*4.1. Experimental work*

The reflection system and the transmission system both comprise a display screen (Philips 246V6QSB, 1920×1080 pixels, 0.272 $mm$ pixel spacing in both directions) and a single camera (AVT Manta G-917B, 25 $mm$ focal length, 3384×2710 pixels, 5.5 $\mu m$ pixel spacing).

A glass plate with ten scratches is used as the detected sample in subsequent experiments, and the width of the scratches varies from 5 to 50 $\mu m$, with the length 5 $mm$ and the depth 2 $\mu m$. The difference in width between two adjacent scratches is 5 $\mu m$.

Similar to the simulation items in Section 3, the experiments in this section are conducted from four items based on the reflection system and transmission system, and the detected results are shown in Figs. 5(a)-5(d) and 6(a)-6(d). The grayscale of these figures has been stretched to 0-255. For better comparison of these detected results, the one-dimensional results in $x$ direction extracted from Figs. 5(a)-5(d) and Figs. 6(a)-6(d) are placed in Fig. 5(e) and Fig. 6(e), respectivly. To observe the peaks and valleys more clearly, a 10-pixel movement has been performed among different detection results.

From Fig. 5 and Fig. 6, it can be discovered that, whether it is the modualtion detection or the direct detection under type 1 binary structured light, scratches will lead to a decline of gray values at the places where they are situated. The image contrast of modulation is better than the contrast obtained under type 1 binary structured light, which is consistent with the simulation results in Figs. 4(b)-4(c). Meanwhile, the scratches captured under uniform illumination are almost invisible, which also conform to the results in Figs. 4(b)-4(c). As for the direct detection based on type 2 binary structured light, the gray values of the scratches show a significant rise, which is the same as the simulation result in Fig. 4(d). Moreover, the peak-to-valley value (PV) of the images captured under type 2 binary structued light is almost equal to the PV value in modulation results, which is confirmed by Figs. 5(e) and 6(e).

In general, the results captured under type 1 or type 2 binary structured light have solved the issue of invisibility of contaminants and defects in uniform illumination system. According to PV values, the results captured under type 2 binary structured light are nearly as good as modulation results. Furthermore, the modulation calculation process is avoided and the number of frames taken by the camera is greatly reduced in the direct detection system, which means that the DSIT is much faster than the modulation detection method.

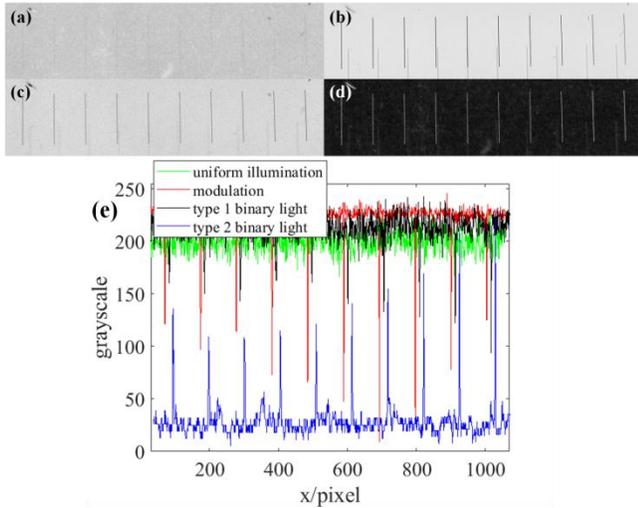

Fig. 5. In the reflection system, the result obtained based on (a) uniform illumination, (b) modulation, (c) type 1 binary structured light, (d) type 2 binary structured light. (e) The one-dimensional results in $x$ direction extracted from (a)-(d).

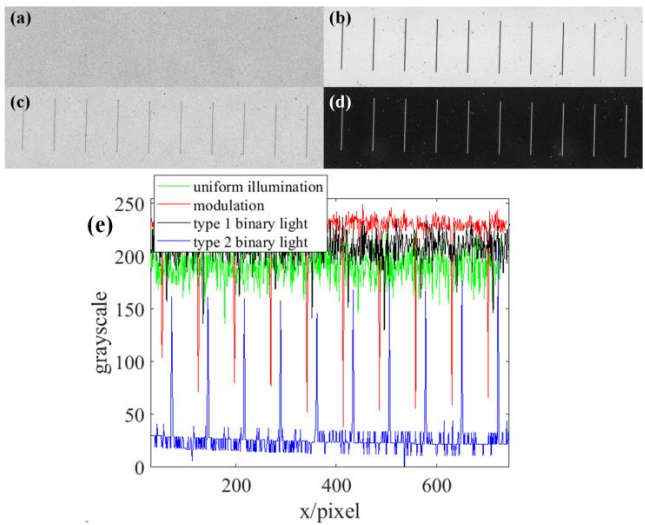

Fig. 6. In the transmission system, the result obtained based on (a) uniform illumination, (b) modulation, (c) type 1 binary structured light, (d) type 2 binary structured light. (e) The one-dimensional results in $x$ direction extracted from (a)-(d).

*4.2. Characteristics of DSIT*

The DSIT proposed in this paper is implemented based on two types of binary structured light. According to the mechanism of contaminant and defect detection, some other kinds of binary structured light can also be used, such as colored binary structured light, which will not be discussed in detail here.

Besides the merits of DSIT summarized by comparing with the luminous flux results under uniform illumination and the modulation results, DSIT has some other prominent features that make it suitable for optical component inspection in industrial environments. First, the illumination based on structured light solves the issue of uneven light distribution in many other detection systems [2]. Second, DSIT inherits the advantage of large detection field of conventional structured light systems [2]. Third, due to the optical path, DSIT based on the transmission system eliminates the background interference caused by the loading platform and is applicable for the detection of transparent objects with large curvature [2]. Fourth, DSIT based on the transmission system overcomes the shortcoming that conventional backlight systems can only find opaque defects [2], which can be used to search many surface defects such as scratches. In addition, DSIT avoids the appearance of residual ripples in general structured light detection system [11].

All these features make the images obtained through DSIT stable and less disturbed, and the captured images can be easily processed and recognized, which is friendly to commercial applications.

## 5. CONCLUSION

In this work, a novel Direct Structured-Light Inspection Technique is proposed for contaminant and defect detection of specular surfaces and transparent objects. DSIT has solved the issue of invisibility of contaminants and defects in uniform illumination system, and according to PV values, the detection effect attained under type 2 binary structured light is comparable to the results calculated based on SMAT. At the same time, due to the difference in detection mechanism, the operation speed of DSIT is much faster than conventional structured light detection methods. In addition, compared with many other inspection methods, DSIT has some other strengths such as large detection field and high image quality, which makes it suitable for contaminant and defect detection in industrial environments.

**Disclosures:** The author declares no conflicts of interest.